\documentclass[pre,twocolumn,showpacs,amsfonts, amsmath,amssymb,superscriptaddress]{revtex4} 
\usepackage{hyperref}
\usepackage{graphicx}			
\usepackage{color}
\graphicspath{{Figures/},.}

\newcommand{\etal}{{\textit{et~al.}}}

\begin{document}
\title{Robustness of Trans-European Gas Networks}

\author{Rui Carvalho}
\email{rui@maths.qmul.ac.uk}
\affiliation{School of Mathematical Sciences, Queen Mary University of London, 
Mile End Road, London E1 4NS, U.K.}

\author{Lubos Buzna}
\email{lbuzna@ethz.ch}
\affiliation{ETH Zurich, UNO C 14, Universit\"atstrasse 41, 8092 Zurich, Switzerland}
\affiliation{University of Zilina, Univerzitna 8215/5, 01026 Zilina, Slovakia}

\author{Flavio Bono}
\email{flavio.bono@jrc.it}
\affiliation{European Laboratory for Structural Assessment,
Joint Research Centre,
Institute for the Protection and Security of the Citizen (IPSC),
Via. E. Fermi, 1 TP 480, Ispra 21027 (VA), Italy}

\author{Eugenio Guti\'errez}
\affiliation{European Laboratory for Structural Assessment,
Joint Research Centre,
Institute for the Protection and Security of the Citizen (IPSC),
Via. E. Fermi, 1 TP 480, Ispra 21027 (VA), Italy}

\author{Wolfram Just}
\affiliation{School of Mathematical Sciences, Queen Mary University of London, 
Mile End Road, London E1 4NS, U.K.}

\author{David Arrowsmith}
\affiliation{School of Mathematical Sciences, Queen Mary University of London, 
Mile End Road, London E1 4NS, U.K.}

\begin{abstract}
Here we uncover the load and fault-tolerant backbones of the trans-European gas pipeline network. 
Combining  topological data with information on inter-country flows, we estimate the global load of the network and its tolerance to failures. To do this, we apply two complementary methods generalized from the betweenness centrality and the maximum flow. We find that the gas pipeline network has grown to satisfy a dual-purpose: on one hand, the major pipelines are crossed by a large number of shortest paths thereby increasing the efficiency of the network; on the other hand, a non-operational pipeline causes only a minimal impact on network capacity, implying that the network is error-tolerant. These findings suggest that the trans-European gas pipeline network is robust, i.e., error tolerant to failures of high load links.
\end{abstract}
\maketitle 
\section{Introduction}
\label{sec:intro}
The world is going through a period when research in energy is overarching \cite{APS08, NatureEditorialJuly08}. Oil and gas prices are volatile because of geopolitical and financial crises. The rate of world-wide energy consumption  has been accelerating, while gas resources are dwindling fast. Concerns about national supply and security of energy are on top of the political agenda, and global climate changes are now believed to be caused by the release of greenhouse gases into the atmosphere~\cite{APS08}. 

Although physicists have recently made substantial progress in the understanding of electrical power-grids~\cite{Watts98, Sachtjen00, Motter02, Albert04, Kinney05, Rosato07, Buzna08, Sole08}, surprisingly little attention has been paid to the structure of gas pipeline networks. Yet, natural gas is often the
energy of choice for home heating and it is increasingly being
used instead of oil for transportation~\cite{NatureEditorialJuly08}, ~\cite{eurostat}. Although renewable energy sources offer the best cuts in overall $CO_2$ emissions, the generation of electricity from natural gas instead of coal can significantly reduce the release of carbon dioxide to the atmosphere.
As the demand for natural gas rises in Europe, it becomes more important to gain insights into the global transportation properties of the European gas network. Unlike electricity, with virtually instantaneous transmission, the time taken for natural gas to cross Europe is measured in days. This implies that the coordination among transport operators is less critical than for power grids. Therefore, commercial interests of competing operators often lead to incomplete or incorrect network information, even at the topological level. Until now, 
modelling has typically been made in small systems by the respective operators, who have detailed knowledge of their own infrastructures. Nevertheless, Ukraine alone transits approximately $80\%$ of Russian gas exports to Europe~\cite{Reymond07}, suggesting the presence of a strong transportation backbone crossing several European countries.     

Historically, critical infrastructure networks have evolved under the pressure to minimize local rather than global failures~\cite{Kim07}. However, little is known on how this local optimization impacts network robustness and security of supply on a global scale. The failure of a few important links may cause major disruption to supply in the network, not because these links connect to degree hubs, but because they are part of major transportation routes that are critical to the operation of the whole network. Here we adopt the view that a \textit{robust infrastructure network} is one which has evolved to be error tolerant to failures of high load links. Our method is slightly different from previous work on real world critical infrastructure networks with percolation theory~\cite{Albert04, Crucitti04, Motter04, Sole08}, which assume the simultaneous loss of many unrelated network components. The absence of historical records on the simultaneous failure of a significant percentage of components in natural gas networks implies that the methods of percolation theory are of little practical relevance in our case. 
Hence, our approach to the challenge of characterising the robustness of global transport on the European gas network was to characterise the \textit{hot transportation backbone} which emerges when measuring network load and error tolerance.

\section{Trans-European Gas Networks}
\label{sec:data sets}

\begin{table}[phtb]\
\caption{\label{tab:statistics}Basic network statistics for the transmission and complete European gas pipeline networks. The complete network is the union of the transmission and distribution networks.}
\begin{ruledtabular}
\begin{tabular}{lll}
{\bf Statistics} & {\bf Gas network}& {\bf Gas network} \\
& {\bf (transmission)} & {\bf (complete network)}\\
\hline
Number of nodes & 2207& 24010\\
Number of edges & 2696 & 25554\\
Total length [km] &119961 & 436289\\
\end{tabular}
\end{ruledtabular}
\end{table}

\begin{figure*}[phtb]
\includegraphics[width=1\textwidth]{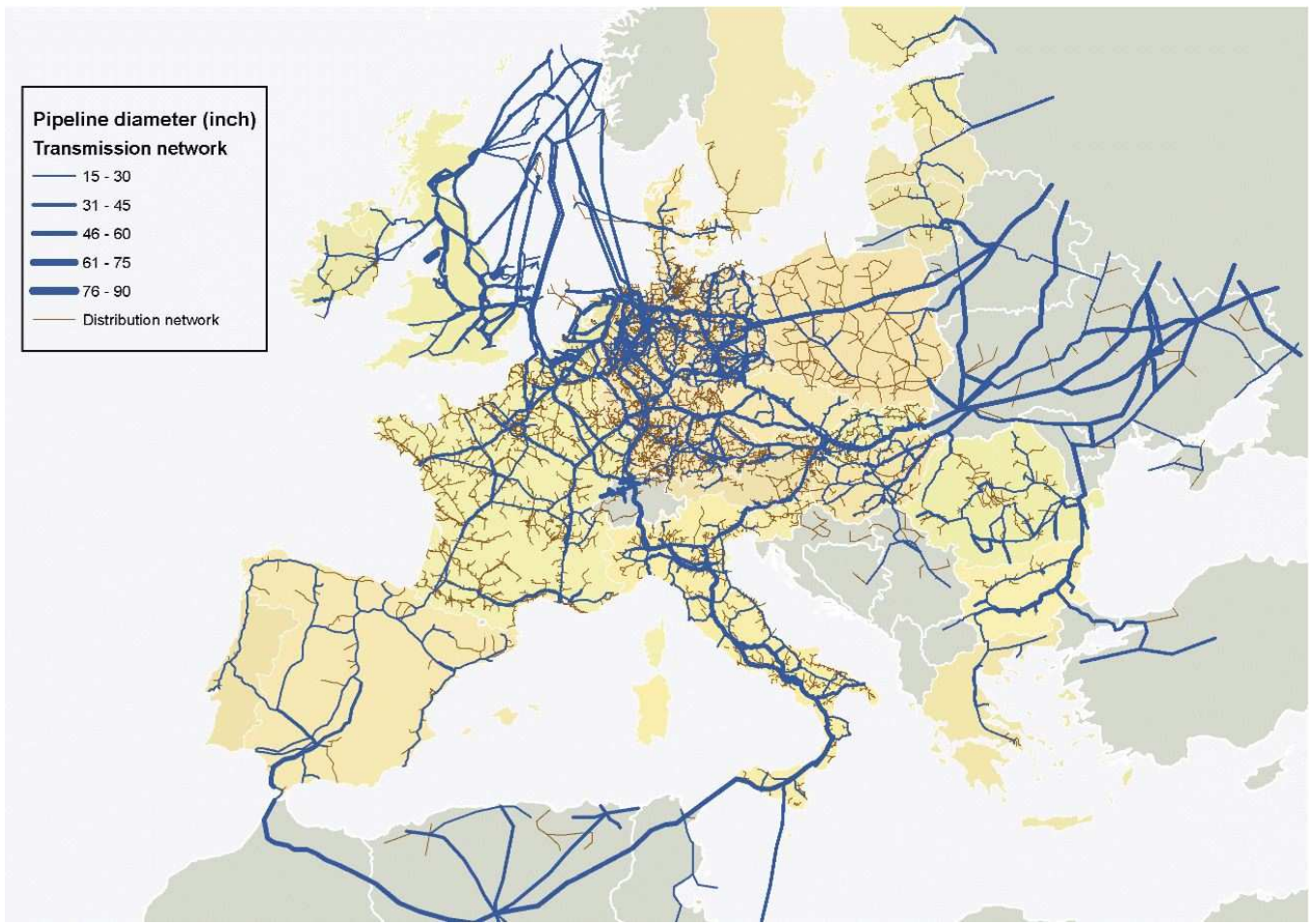}
\caption{\label{fig:EU_Gas_Net}(Colour online) European gas pipeline network. We show the transmission network [blue (dark gray) pipelines] overlaid with the distribution network [brown (light gray) pipelines]. Link thickness is proportional to the pipeline diameter. We projected the data with the Lambert azimuthal equal area projection~\cite{Robinson95}. Background colours identify EU member states.}
\end{figure*}

\begin{figure}[phtb]
\includegraphics[width=0.5\textwidth]{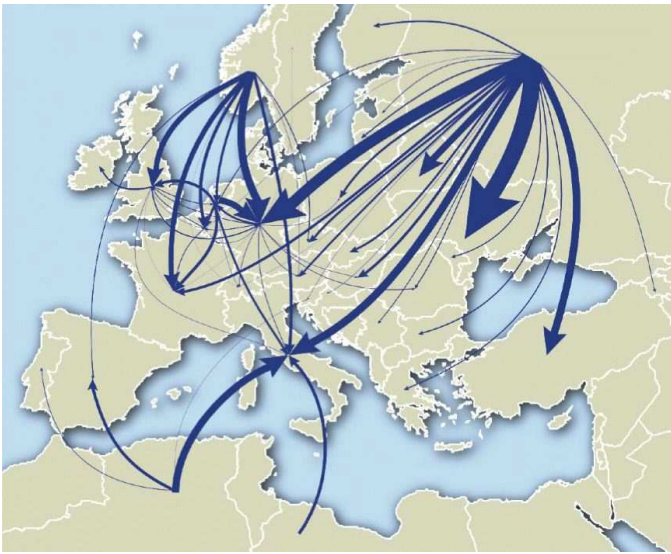}
\caption{\label{fig:trade_network}(Colour online) Network of international gas trade movements by pipeline~\cite{BP08}. Link thickness is proportional to the annual volume of gas traded.}
\end{figure}

\begin{figure}[phtb]
\includegraphics[width=0.5\textwidth]{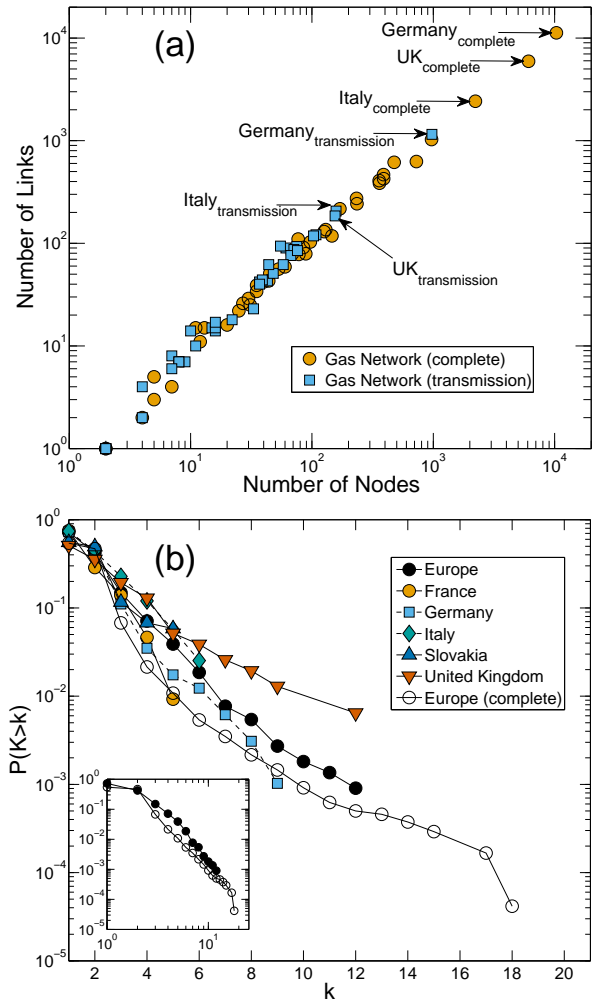}
\caption{\label{fig:avg_node_degree}(Colour online)  (a) Plot of the number $L$ of links versus the number $N$ of nodes for the transmission and complete (i.e. transmission and distribution) natural gas networks of the countries analysed. The three largest national transmission and complete networks are also labeled (Germany, UK and Italy). (b) Plot of the complementary cumulative degree distribution of the European gas networks, together with national transmission gas networks larger than $100$ pipelines. The inset shows the degree distribution of the European complete and transmission networks on a double-logarithmic scale,  highlighting the presence of fat tails on the degree distribution of the complete gas pipeline network.}
\end{figure}

We have extracted the European gas pipeline network from the Platts Natural Gas geospatial data \cite{platts}. The data set cover all European countries (including non EU countries such as Norway and Switzerland), North Africa (main pipelines from Morocco and Tunisia), Eastern Europe (Belarus, Ukraine, Lithuania, Latvia, Estonia, and Turkey) and Western Russia (see Fig. \ref{fig:EU_Gas_Net}).

Similarly to electrical power grids, gas pipeline networks have two main layers: transmission and distribution. The transmission network transports natural gas over long distances (typically across different countries), whereas pipelines at the distribution level cover urban areas and deliver  gas directly to end consumers. We extracted the gas pipeline transmission network
from the complete natural gas network, as the connected component composed of all the important pipelines with diameter $d \geq 15$ inches. To finalize the network, we added all other pipelines interconnecting major branches \cite{gte}. We treated the resulting network as undirected due to the lack of information on the direction of flows. However, network links are weighted according to pipeline diameter and length. 

The European gas pipeline infrastructure is a continent-wide sparse network which crosses $38$ countries, has about $2.4\times10^4$ nodes [compressor stations, city gate stations, liquefied  natural gas (LNG) terminals, storage facilities, etc.] connected by approximately $2.5\times10^4$ pipelines (including urban pipelines), spanning more than $4.3\times10^5$ km (see Table~\ref{tab:statistics}). 
The trans-European gas pipeline network is, in fact, a union of national infrastructure networks for the transport and delivery of natural gas over Europe. These networks have grown under different historical, political, economic, technological and geographical constraints, and might be very different from each other from a topological point of view. The Platts data set did not include volume or directionality of flows. Hence, we assessed the global structure of the European gas network under the availability of incomplete information on flows. To reduce uncertainty on flows, we combined the physical infrastructure network with the network of international natural gas trade movements by pipeline for 2007~\cite{BP08} (see Fig.~\ref{fig:trade_network}).

To investigate similarities among the national gas networks, we first plotted in Fig.~\ref{fig:avg_node_degree}(a) the number $N$ of nodes, versus the number $L$ of links for each country. Figure~\ref{fig:avg_node_degree}(a) suggests that both the transmission and the complete (i.e., transmission and distribution) networks have approximately the same average degree because all points fall approximately along a straight line. Indeed, we found $\left\langle k_{\text{transmission}}\right\rangle = 2.4$ and $\left\langle k_{\text{complete}}\right\rangle =2.1$~\footnote{Ricard Sole~\etal~found similar results for European power grids~\cite{Sole08}.}. Surprisingly, the size of the complete European national gas networks ranges over three orders of magnitude from two nodes (former Yugoslav Republic of Macedonia) to $10334$ nodes (Germany). Further, the German transmission network is considerably larger than the Italian network. Germany  has a long history of industrial usage of gas and is a major hub for imports from Russia and the North Sea~\cite{eia, eurostat}.

Given that the national networks have very different sizes, but approximately the same average degree, we looked for regularities in the probability distribution of degree of the European gas networks [see Fig.~\ref{fig:avg_node_degree}(b)]. In accordance with previous studies of electrical transmission networks~\cite{Amaral00,Sole08}, the complementary cumulative degree distribution of the transmission network decays exponentially as $P(K>k)\approx \exp (-k/\lambda )$, with $\lambda =1.44$.
Unexpectedly, we found that the degree distribution of the complete gas network is heavy tailed, as can be seen in the inset of Fig.~\ref{fig:avg_node_degree}(b). This suggests that the gas network may be approximated by an exponential network at transmission level, but not when the distribution level is considered as well. The distribution network is mainly composed of trees which attach to nodes in the transmission network, thus forming the complete network. Hence, the fat tails are a combined effect of increasing the degree of existing transmission nodes in the complete network, and adding distribution nodes with lower degrees.

\begin{figure}[phtb]
\includegraphics[width=0.5\textwidth]{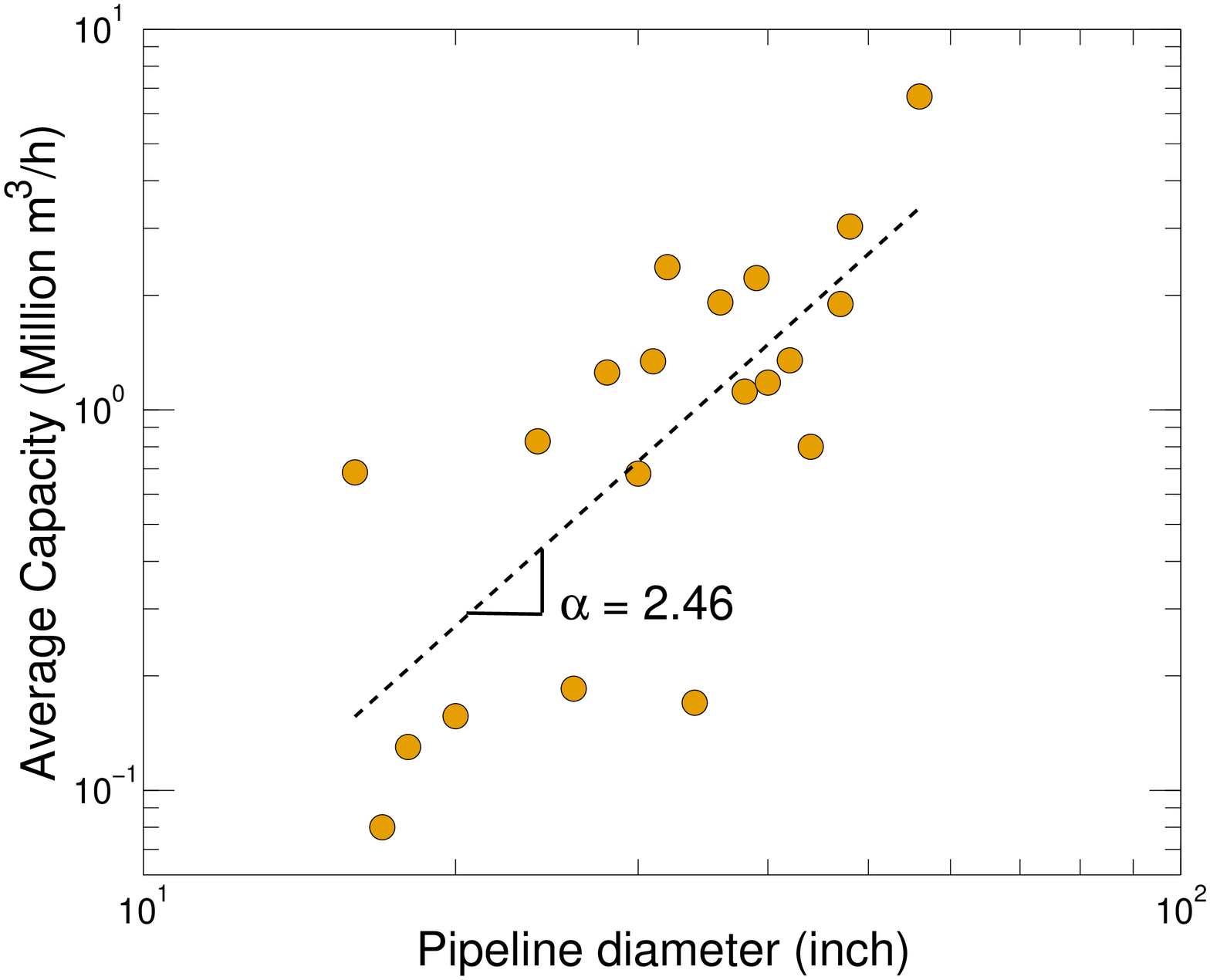}
\caption{\label{fig:GTE_vs_D}(Colour online) Plot of the digitized Gas Transmission Europe (GTE) pipeline capacity versus pipeline diameter on a double logarithmic scale. We digitized the European gas network map from GTE and assigned the GTE capacities to pipelines in the Platts data set. The straight line is a regression to the data, which corresponds to $c=a d^{\alpha}$ with $\alpha=2.46$.}
\end{figure}

\begin{figure}[phtb]
\includegraphics[width=0.5\textwidth]{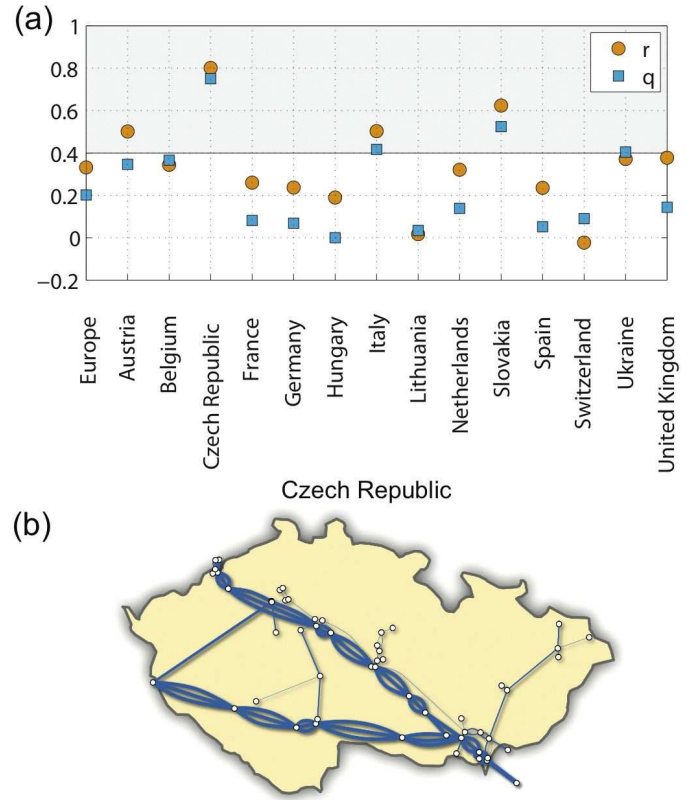}
\caption{\label{fig:Cij_vs_kikj}(Colour online) (a) Pearson correlation coefficient, $r$, for the degrees of two linked nodes and the capacity of the linking pipeline; and percentage of capacity on parallel pipelines, $q$. We considered only countries with more than $50$ pipelines. (b) Network layout for the Czech Republic. Link thickness is proportional to the pipeline capacity. }
\end{figure}

The July 2007 release of the Platts data set, which we analysed, did not include information on  the capacity of pipelines. To estimate the pipeline capacity, we compared cross-border flows based on capacity estimates for an incompressible fluid, where the capacity $c$ can shown to scale as $d^{\gamma}$ with pipeline diameter $d$ (see the Appendix), to reported cross-border flows extracted from the digitized Gas Transmission Europe (GTE) map~\cite{gte}. Figure~\ref{fig:GTE_vs_D} is a plot of averaged pipeline capacity versus pipeline diameter in a double-logarithmic scale for pipelines in the GTE data set. We found a good match between the theoretical prediction of $c\sim d^{\gamma}$ with $\gamma \simeq 2.6$, and the capacity of major pipelines as made evident by the regression to the data $c\sim d^{\alpha}$ with $\alpha=2.46$. Hence, we used the exponent $\gamma=2.5$ as a trade-off between the theoretical prediction and the numerical regression, and approximated $c\sim d^{2.5}$.

To understand the national structure of the network, we investigated the tendency of highly connected nodes to link to each other over high capacity pipelines. Figure~\ref{fig:Cij_vs_kikj}(a) shows the Pearson correlation coefficient between the product of the degrees of two nodes connected by a pipeline and the capacity of the pipeline, 
\begin{equation}
r=\frac{\sum_{e_{ij}}\left(  k_{i}k_{j}-\overline{k_{i}k_{j}}\right)  \left(
c_{e_{ij}}-\overline{c_{e_{ij}}}\right)  }{\sqrt{\sum_{e_{ij}}\left(
k_{i}k_{j}-\overline{k_{i}k_{j}}\right)  ^{2}}\sqrt{\sum_{e_{ij}}\left(
c_{e_{ij}}-\overline{c_{e_{ij}}}\right)  ^{2}}}
\end{equation}
where $k_i$ and $k_j$ are the degrees of the nodes at the ends of pipeline $e_{ij}$, and $c_{e_{ij}}$ is the capacity of the pipeline. Countries with high values of $r$ have a gas pipeline network where degree hubs are interconnected by several parallel pipelines. Further, we plotted the percentage $q$ of capacity on parallel pipelines for each national network. Typically, countries with high values of $r$ also have high values of $q$.

Figure~\ref{fig:Cij_vs_kikj}(a) shows relatively large differences among countries: Austria, the Czech Republic, Italy and Slovakia have both high values of $r$ and $q$. A visual inspection of the network in these countries uncovers the presence of many parallel pipelines organized along high capacity corridors [see Fig.~\ref{fig:Cij_vs_kikj}(b)]. 
Taken together, these results suggest that some European national networks have grown structures characterised by chains of high capacity (parallel) pipelines over-bridging long distances. This has the consequence of improving the local error tolerance because the failure of one pipeline implies only a decrease in flow.

Motivated by the finding of error tolerance in the gas pipeline networks, we then asked the question of whether there are global topological properties of the European network which could characterize the network robustness.

\section{Analysis of man-made distribution networks with incomplete flow information}
\label{sec:Methods}

To gain insights into the overlaid infrastructure and aggregate flow networks, we propose two complementary approaches which aim at identifying the backbones of the overlaid networks in Figs.~\ref{fig:EU_Gas_Net} and ~\ref{fig:trade_network}. In both approaches, the flow network allows an approximate estimate of the volume of directed flows as we will detail in \ref{sec:BC_in_Fn}.

In the first approach, we assume that transport occurs along the shortest paths in geographical space. We search for a global backbone characterised by the presence of flow corridors where individual components were designed to sustain high loads.

In the second approach, we look at fault tolerance when single components fail. Recent studies of network vulnerability in infrastructure networks suggest that, although these networks have exponential degree distributions, under random errors or attacks the size of the percolation cluster decreases in a way which is reminiscent of scale-free networks~\cite{Rosas-Casals07,Sole08}. Typically, these studies presume that  a large percentage of nodes or links may become nonoperational simultaneously, i.e., the time scale of node or link failure is much faster than the time scale of repair. Whereas the underlying scenario of a hacker or terrorist attack on an infrastructure network causing large damage is certainly worth studying~\cite{Williams07}, the consequences of such attacks may not be assessed properly when measuring damage by the relative size of the largest component. 
Here, we estimate the loss of flow when a single link is non-operational. We search for a global backbone characterised by corridors of interconnected nodes, where the removal of one single link causes a high loss of flow from source to sink nodes. 

The maximum flow problem and the shortest paths problem are complementary, as they capture different aspects of minimum cost flow. Shortest path problems
capture link lengths, but not capacities; maximum flow problems model
link capacities, but not lengths.

\subsection{Generalized betweenness centrality}
\label{sec:BC_in_Fn}

Many networks are in fact substrates, where goods, products, substances or materials flow from sinks to sources through components laid out heterogeneously in geographical space. Examples range from supply networks~\cite{Helbing04}, spatial distribution networks~\cite{Gastner06} and energy networks~\cite{Albert04} to communication networks~\cite{Mukherjee08}. Node and link stress in these networks is often characterised by the betweenness centrality. Consider a substrate network $G_S=(V_S,E_S)$ with node-set $V_S$ and link-set $E_S$. The betweenness centrality of link $e_{ij}\in E_S$ is defined as the relative number of shortest paths between all pairs of nodes which pass through $e_{ij}$,
\begin{equation}
g(e_{ij})=\sum_{\substack{ s,t\in V_S \\ s\neq t}}\frac{\sigma _{s,t}(e_{ij})}{%
\sigma _{s,t}}  \label{eq:betweenness_centrality}
\end{equation}
where $\sigma _{s,t}$ is the number of shortest paths from node $s$ to node $%
t$ and $\sigma _{s,t}(e_{ij})$ is the number of these paths passing through
link $e_{ij}$. The concept of betweenness centrality was originally developed to characterise the influence of nodes in social networks~\cite{Freeman77,Koschutzki05} and, to our knowledge, was used for the first time in the physics literature in the context of social networks by Newman~\cite{Newman01} and in the context of communication networks by Goh~\etal~\cite{Goh01}\footnote{Goh~\etal~define load similarly to betweenness centrality, but the two measures are different in the presence of more than one shortest path between two nodes.}. 

Betweenness centrality is relevant in man-made networks which deliver products, substances or materials as cost constraints on these networks condition transportation to occur along shortest paths. However, nodes and links with high betweenness in spatial networks are often near the network barycentre~\cite{Barrat05}, whose location is given by $\mathbf{x}_{G}=\sum_{i}x_{i}/N$,
whereas the most important infrastructure elements are frequently along the periphery, close to either the sources or the sinks. Although flows are conditioned by a specific set of sources and sinks, the traffic between these nodes may be highly heterogeneous and one may have only access to aggregate transport data, but not to the detailed flows between individual sources and sinks (e.g. competition between operators may prevent the release of detailed data). Here we propose a generalization of betweenness centrality in the context of flows taking place on a substrate network, but where flow data are available only at an aggregate level. We then show in the next section how the generalized betweenness centrality can help us to gain insights into the structure of trans-European gas pipeline networks.

The substrate network is often composed of sets of nodes which act like aggregate sources and sinks. The aggregation can be geographical (e.g., countries, regions or cities), or organizational (e.g., companies or institutions). If the flow information is only available at aggregate level then a possible extension of the betweenness centrality for these networks is to weight the number of shortest paths between pairs of source and sink nodes by the amount of flow which is known to go through the network between aggregated pairs of sources and sinks. To do this, we must first create a flow network by 
partitioning the substrate network, $G_S=(V_S,E_S)$, into a set of disjoint subgraphs $V_F=\{(V_{S_{1}},E_{S_{1}}),\cdots ,(V_{S_{M}},E_{M})\}$. The flow network $G_F=(V_F, E_F)$ is then defined as the directed network of flows among the subgraphs in $V_F$, where the links $E_F$ are weighted by the value of aggregate flow among the $V_F$. For our purposes, the substrate network is the trans-European gas pipeline network represented in Fig.~\ref{fig:EU_Gas_Net} and the flow network is the network of international gas trade movements by pipeline in Fig.~\ref{fig:trade_network}.
  
The \textit{generalized betweenness centrality} (generalized betweenness) of link $e_{ij}\in E_S$ is defined as follows. Let $T_{K,L}$ be the flow from source subgraph $K=(V_K,E_K)\in V_F$ to sink subgraph $L=(V_L,E_L)\in V_F$. Take each link $e_{KL}\in E_F$ and compute the betweenness centrality from Eq. (\ref{eq:betweenness_centrality})
 of $e_{ij}\in E_S$ restricted to source nodes $s\in V_K$
and sink nodes $t\in V_L$. 
The
contribution of that flow network link is then weighted by $T_{K,L}$ and normalized by the number of links in a complete bipartite graph between nodes in $V_K$ and $V_L$,
\begin{equation}
G_{ij}=\sum_{e_{KL}\in E_F}\sum_{s\in V_K,t\in V_L}\frac{T_{K,L}}{%
|V_K||V_L|}\frac{\sigma _{st}\left( e_{ij}\right) }{\sigma _{st}}.
\label{eq:rb}
\end{equation}

\subsection{Generalized max-flow betweenness vitality}
\label{sec:MF_in_Fn}

The maximum flow problem can be stated as follows: In a network with link capacities,
we wish to send as much flow as possible between two particular nodes, a source and a sink, without exceeding the capacity of any link~\cite{Ahuja93}. Formally, an \textit{s-t flow network} $G_F=(V_F,E_F,s,t,c)$ is a digraph\footnote{To transform the undirected network to a directed network, we replace each undirected link between nodes $i$ and $j$ with two directed links $e_{ij}$ and $e_{ji}$, both with capacity $c_{ij}$.} with node-set $V_F$, link-set $E_F$, two distinguished nodes, a \textit{source s} and a \textit{sink t}, and a capacity function $c:E_F\rightarrow \mathbb{R}_0^+$. A \textit{feasible flow} is a function $f:E_F\rightarrow \mathbb{R}_0^+$ satisfying the following two conditions: $0\leq f(e_{ij}) \leq c(e_{ij}) ,$ $\forall e_{ij}\in E_F$ (capacity constraints)
; and $\sum_{j:e_{ji}\in E_F}f\left( e_{ji}\right) =\sum_{j:e_{ij}\in E_F}f(e_{ij}) ,
$ $\forall i\in V\backslash \{s,t\}$ (flow conservation constraints). The \textit{maximum s-t flow} is defined as the maximum flow into the sink, $%
f_{st}(G_F)=\max (\sum_{i:e_{it}\in E_F}f(e_{it}))$ subject to the conditions that
the flow is feasible~\cite{Ahuja93,Frank71,Jungnickel07}. 

\begin{figure}[phtb]
\includegraphics[width=0.5\textwidth]{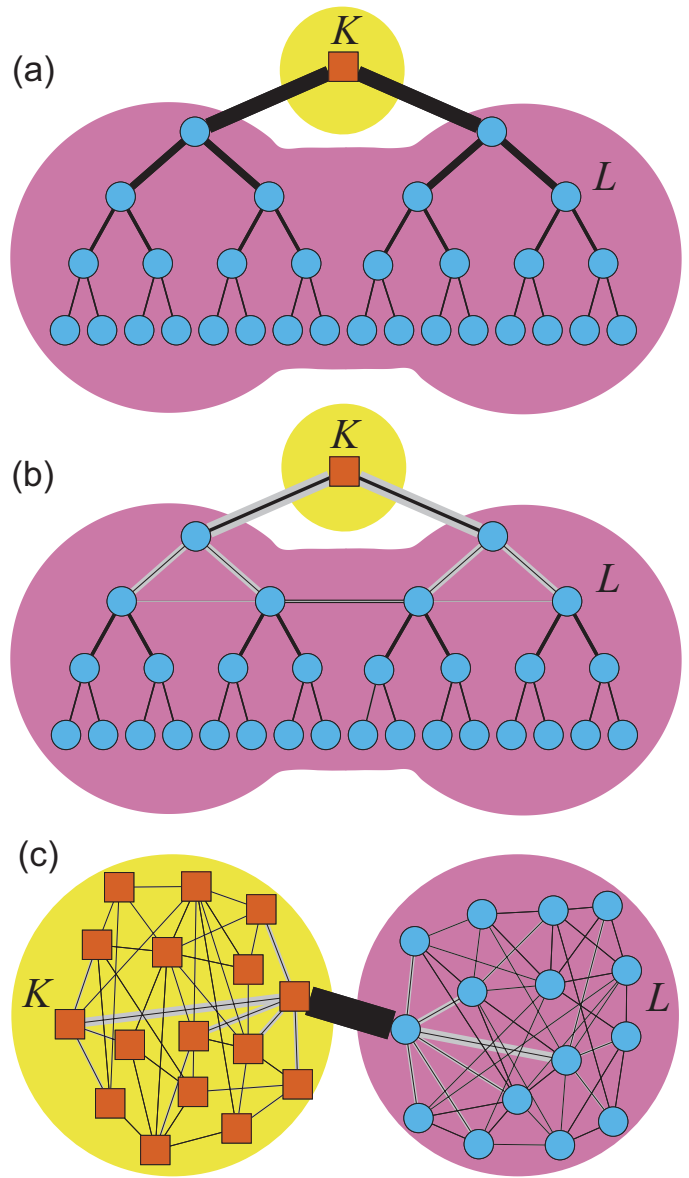}
\caption{\label{fig:Example}(Colour online) Generalized betweenness (gray) and vitality (black) measures on (a) a rooted tree, (b) a modified rooted tree with interconnections at a chosen level, and (c) two communities connected by one link. Nodes are shaped according to their function: source nodes are squares, and sink nodes are circles. Both generalized betweenness and vitality depend on $T_{K,L}$, which is a constant for all examples. The smaller value of the two quantities is always drawn on the foreground so that both measures are visible.}
\end{figure}

\begin{figure*}[phtb]
\includegraphics[width=1\textwidth]{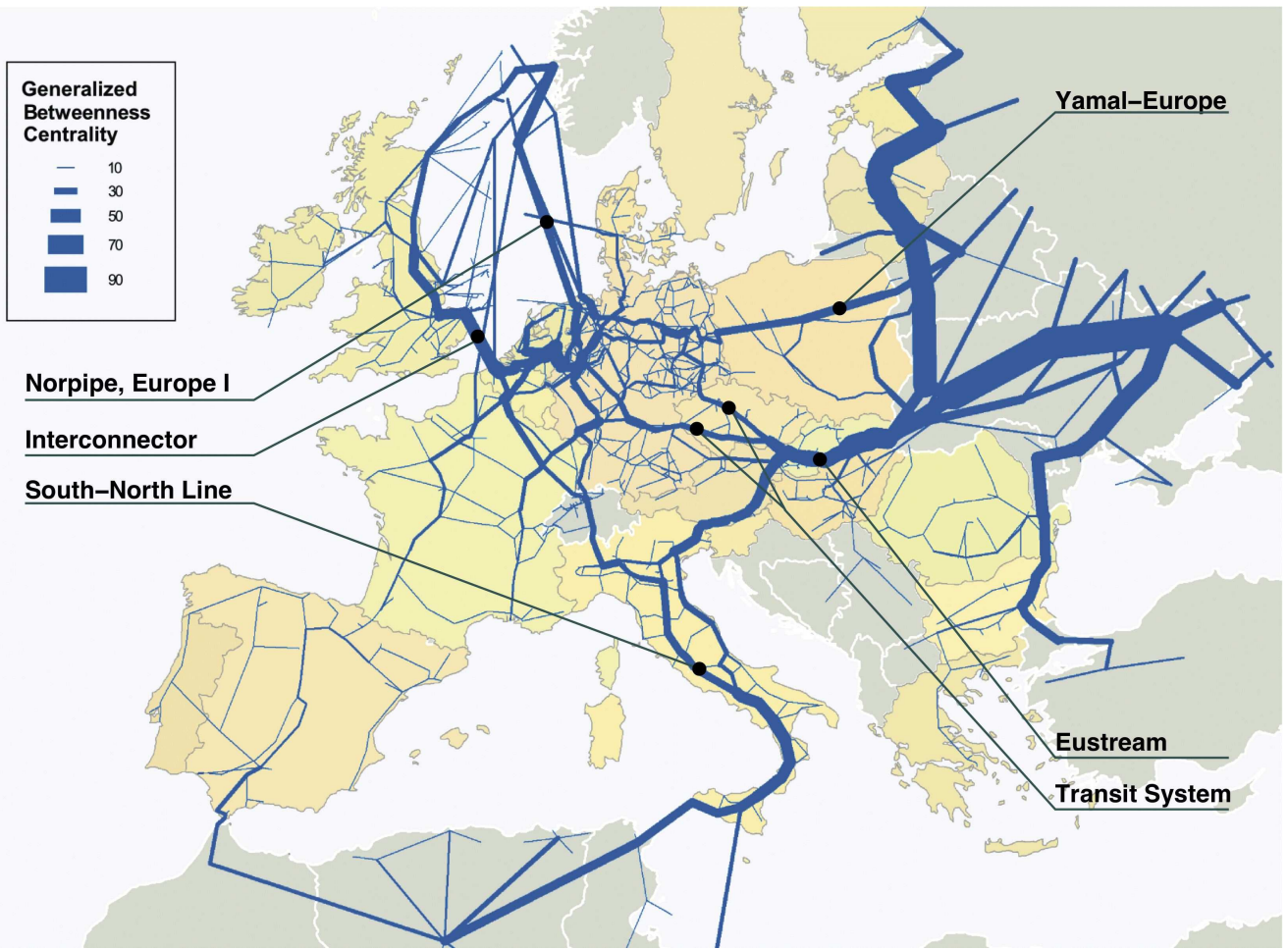}
\caption{\label{fig:Restricted_BC}(Colour online) Trans-European natural gas network. Link thickness is proportional to the generalized betweenness centrality [see Eq.~(\ref{eq:rb}, where the sets $K$ and $L$ are countries and the values of $T_{K,L}$ are taken from the data in Fig.~\ref{fig:trade_network}]. We labeled several major EU pipeline connections. The large difference between the generalized betweenness on these pipelines and the rest of the network suggests that the network has grown, to some extent, to transport natural gas with minimal losses along the shortest available routes.}
\end{figure*}

\begin{figure*}[phtb]
\includegraphics[width=1\textwidth]{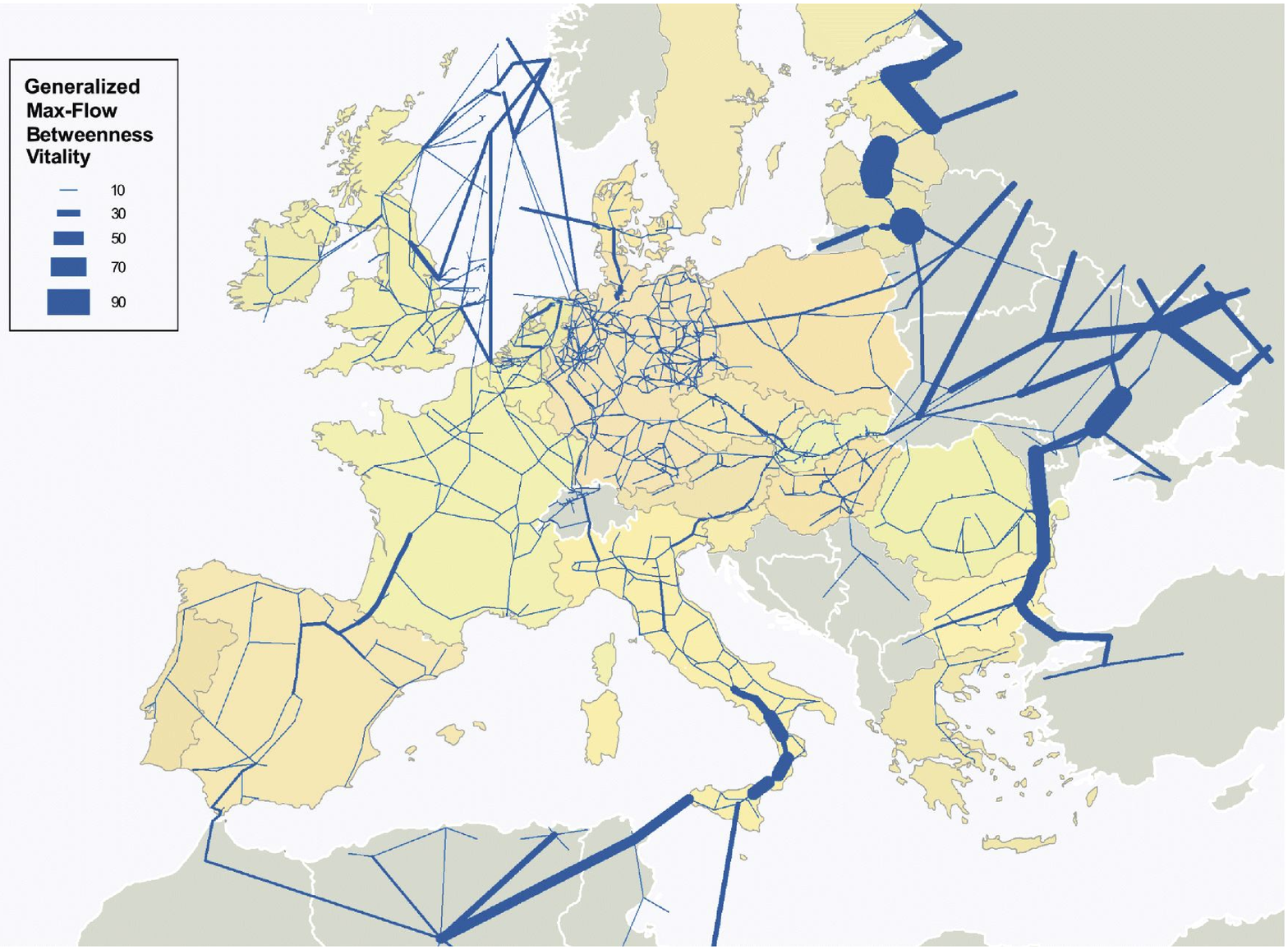}
\caption{\label{fig:Max-Flow-Betweenness} (Colour online) trans-European natural gas network. Link thickness is proportional to the generalized max-flow betweenness vitality [see Eq.~(\ref{eq:max-flow-vitality}), where the sets $K$ and $L$ are countries and the values of $T_{K,L}$ are taken from the data in Fig.~\ref{fig:trade_network}]. Pipelines close to the major sources tend to have high values of generalized vitality because this is where the network bottleneck is located. Pipelines along sparse interconnections between larger parts of the network (e.g., the Spanish-French border) also tend to have high values of generalized vitality, when compared to neighbouring pipelines.}
\end{figure*}

We are now interested in the answer to the question: How does the maximum flow
between all sources and sinks change, if we remove a link $e_{ij}$ from the
network? In the absence of a detailed flow model, we calculated the flow that is lost when a link $e_{ij}$ becomes nonoperational assuming that the network is working at maximum capacity. In agreement
with Eq.~(\ref{eq:rb}), we define the \textit{generalized max-flow betweenness
vitality}~\cite{Koschutzki05, Latora05} (generalized vitality): 
\begin{equation}
V_{ij}=\sum_{e_{KL}\in E_F}\sum_{s\in V_K,t\in V_L}\frac{T_{K,L}}{|V_K||V_L|}\frac{\Delta ^{G_F}_{st}(e_{ij})}{f_{st}(G_F)},  
\label{eq:max-flow-vitality}
\end{equation}
where the amount of flow which must go through link $e_{ij}$ when the
network is operating at maximum capacity is given by the vitality of the
link~\cite{Koschutzki05}: $\Delta ^{G_F}_{st}(e_{ij}) =f_{st}(G_F)
-f_{st}(G_F\backslash {e_{ij}} ) $, and $f_{st}(G_F)$ is
the maximum $s$-$t$ flow in $G_F$. 
\subsection{Generalized Betweenness Centrality versus Max-flow Betweenness Vitality}
\label{sec:Comparison}

A close inspection of Eq.~(\ref{eq:rb}), generalized betweenness, and Eq.~(\ref{eq:max-flow-vitality}), generalized vitality, reveals that both measures have the physical units of gas flow given by $T_{K, L}$. Further, the relative number of shortest paths crossing a link $e_{ij}$ is bounded by $ 0\leq \frac{\sigma _{st}\left( e_{ij}\right) }{\sigma _{st}}\leq 1$, and the relative quantity of flow which must go through the same link $e_{ij}$ is also bounded by  $0\leq \frac{\Delta \left( e_{ij}\right) }{f_{st}}\leq 1$. Thus, the generalized betweenness (\ref{eq:rb}) and generalized vitality (\ref{eq:max-flow-vitality}) can be compared for each link.

To examine the relationship between these two quantities, we considered three simplified illustrative networks: a rooted tree where the root is the source node and all other nodes are sinks, the same rooted tree with additional links interconnecting children nodes at a selected level, and two communities of source and sink nodes connected by one single link. We chose these particular examples because they resemble subgraphs which appear frequently on the European gas pipeline network and thus they may help us to gain insights into the structure of the real world network.  

Both the generalized betweenness and vitality have the same values on the links of trees where the root is the source node and the other nodes are sinks. To see this, consider without loss of generality the case when $T_{K,L}=1$. Then, the generalized betweenness of a link is the proportion of sinks reachable (along shortest paths) over the link, and the generalized vitality is the proportion of sinks fed by the link. The two quantities have the same value on the links of a tree and we illustrated this in Fig.~\ref{fig:Example}(a), where we drew link thickness proportional to the generalized betweenness (gray) and generalized vitality (black).  

Figure~\ref{fig:Example}(b) shows a modified tree network where we have connected child nodes at a chosen level. Here, the shortest paths between the root and any other node are unchanged from the example of the tree, but removing a link $e_{ij}$ situated above the lateral interconnection does not cut all connections between the source (root) and sink nodes. As a consequence the values of generalized vitality are significantly smaller in the upper part of the graphs. Figure~\ref{fig:Example}(c) shows two communities connected by one link, where source nodes are on one community and sink nodes on the other. This example is interesting for two reasons. First, the arguments used to explain why generalized betweenness and vitality take the same values on trees are also valid in this example. Second, the link connecting the two communities has a much higher value of generalized betweenness and vitality than the links inside the communities, which led us to expect that these two measures could hint at the presence of modular structure in the real world network.
\section{Network Robustness}

The generalized betweenness measure, defined by Eq. (\ref{eq:rb}), assumes that gas is transported from sinks to sources along the shortest paths. To investigate whether this hypothesis is correct, we plotted the European gas pipeline network and drew the thickness of each pipeline proportional to the value of its generalized betweenness (see Fig.~\ref{fig:Restricted_BC}). We found that major loads, predicted by the generalized betweenness centrality,  were on the well-known  high capacity transmission interconnections such as the "Transit system" in the Czech Republic, the "Eustream" in Slovakia, the "Yammal-Europe" crossing Belarus and Poland, the "Interconnector" connecting the UK with Belgium or the "Trans-Mediterranean" pipeline linking Algeria to mainland Italy through Tunisia and Sicily. 
The dramatic difference between the values of generalized betweenness of all the major European pipelines and the rest of the network suggests that the network has grown to some extent to transport natural gas with minimal losses along the shortest available routes between the sources and end consumers.
These  major pipelines are the transportation backbone of the European natural gas network. 

During the winter season, cross-border pipelines are used close to their full capacity~\cite{gte}. In this situation, the generalized vitality of a pipeline [Eq. (\ref{eq:max-flow-vitality})] can be interpreted as the network capacity drop, or the amount of flow that cannot be delivered, if that pipeline becomes nonoperational. 
The obvious drawback of the generalized vitality is that it takes into account the overall existing network capacity without considering the length of paths. Conversely, the generalized betweenness considers the length of shortest paths, but not the capacity of pipelines. Since we assess the network from two complementary view points, we expect that the results will allows us to get a more complete picture of the general properties of the European gas pipeline network.

Figure~\ref{fig:Max-Flow-Betweenness} shows the values of the generalized vitality in the European gas pipeline network. 
We found several relatively isolated segments with high a generalized vitality located in Eastern Europe, close to the Spanish-French border, as well as on the south of Italy. The high values of the generalized vitality in Eastern Europe can be explained by two factors. On one hand, the generalized vitality of pipelines close to the sources is higher than elsewhere simply because these pipelines are the bottleneck of the network. On the other hand, our approximation that a directed link in the flow matrix implies gas flowing from all nodes in the source to all nodes in the sink countries was clearly coarse-grained for flows between Russia and the Baltic states, as it would imply that pipelines in southern Russia would also supply the Baltic countries. This highlights boundary effects on the calculation of betweenness vitality, as the data set excluded most of the Russian gas pipeline network. The case of the Spanish-French border was different, though. The link with high vitality separates the Iberian Peninsula from the rest of mainland Europe. If this link was to be  cut, then Portugal and Spain would only be linked to the pipeline network through Morocco. Finally, the south of Italy highlights an interesting example of two communities (Europe and North Africa) separated by the Trans-Mediterranean pipeline, which is reminiscent of the example in Fig.~\ref{fig:Example}(c).

Perhaps surprisingly, we found that the generalized vitality is more or less homogeneous in most of mainland Europe. This result suggests that the EU gas pipeline network has grown to be error tolerant and robust to the loss of single links. 

Distribution networks originate from the need for an effective connectivity among sources and sinks~\cite{Banavar99, Gastner06a}.  For example, a spanning tree is highly efficient as it transports goods from sinks to sources in a way that shortens the total length of the network, thereby increasing its efficiency and viability. If the European gas pipeline network had been built as a spanning tree, its links would have very similar values of generalized betweenness and max-flow vitality (see Fig.~\ref{fig:Example}). 

The values of the generalized betweenness are considerably higher than the corresponding values of generalized vitality for the most important pipelines in the European Union. In other words, the major pipelines are crossed by many shortest paths, but a nonoperational pipeline causes only a minor capacity drop in the network. This dramatic contrast between the two measures reveals a \textit{hot backbone}~\cite{Almaas04} showing that the trans-European gas pipeline network is robust, i.e. error tolerant to failures of high load links.
\section{Conclusions}

We analysed the trans-European gas pipeline network from a topological point of view. We found that the European national gas pipeline networks have approximately the same value of average node degree, even if their sizes vary over three orders of magnitude. Like the electrical power grid, the degree distribution of the European gas transmission network decays exponentially. Unexpectedly, the degree distribution of the complete (transmission and distribution) gas pipeline network is heavy tailed. In some countries which are crucial for the transit of gas in Europe (Austria, the Czech Republic, Italy and Slovakia), we found that the main gas pipelines are organized along high capacity corridors, where capacity is split among two or more pipelines which run in parallel, over-bridging long distances. This implies that the network is error tolerant because the failure of one pipeline causes only a decrease in flow. Motivated by the finding of error tolerance in national networks, we then addressed the problem of capturing the topological structure of the European gas network.

At a global scale, the growth of the European gas pipeline network has been determined by two competing mechanisms. First, the network has grown under cost and efficiency constraints to minimize the length of transport routes and maximize transported volumes. Second, the network has developed error tolerance by adding redundant links. The combination of the two mechanisms guarantees that the European gas pipeline network is robust, i.e. error tolerant to failures of high load links. To reveal the network robustness, we analysed two measures---the generalized betweenness and generalized vitality---which highlight global backbones of transport efficiency and error tolerance, respectively. Finally, we proposed that the hot backbone of the network is the skeleton of major transport routes where the network is robust, in other words, where values of generalized betweenness are high and values of generalized vitality are low.  Our method is of potential interest as it provides a detailed geographical analysis of engineered distribution networks. 

Further research in continent-wide distribution networks could proceed along several directions. The optimality of existing networks and the existence of scaling laws could be approached from a theoretical perspective~\cite{Durand04,Durand06}. 
Planned and under-construction pipelines may change the robustness of the network, in particular within their geographical vicinity. Liquefied  natural gas (LNG), which is nowadays transported at low cost between continents, is increasingly supplying the pipeline network. The combined effect of LNG and storage facilities throughout the European coastline has the potential to reduce the dependency on one single exporting country, such as Russia. Last, but not the least, the dispute between Russia and Ukraine in January 2009 has brought supply security to the top of the European political agenda and highlighted how the European gas network is robust to engineering failures, yet fragile to geopolitical crises. 

\begin{acknowledgments}
We wish to thank Dirk Helbing, Sergi Lozano, Amin Mazloumian and Russel Pride for valuable comments, and gratefully acknowledge the support of EU projects MANMADE (Grant No. 043363) and IRRIIS (Grant No. 027568).
\end{acknowledgments}

\appendix*
\section{Pipeline capacity}
\label{sec:Pipeline_Capacity}

The capacity of a pipeline can be schematically derived as follows: It is known that the flow of an incompressible viscous fluid in a circular pipe can be described in the laminar regime (with a parabolic velocity profile) by the Hagen-Poiseuille equation~\cite{Landau87}, which states that the volume of fluid passing per unit time is 

\begin{equation}
dV/dt=\pi \Delta pr^{4}/\left( 8\eta l\right),
\end{equation}
where $\Delta p$ is the pressure difference between the two ends of the pipeline, $l$ is the length of the pipeline (thus, $-\Delta p/l$ is the pressure gradient), $\eta$ is the dynamic viscosity, and $r$ is the radius of the pipeline. 
However, the gas network operates in the turbulent regime, and the Hagen-Poiseuille equation is no longer valid. Therefore, we apply the  Darcy--Weisbach equation for pipeline head loss, $h_{f}$. This  is a phenomenological equation which describes the loss of energy due to friction within the pipeline and is valid in the laminar and turbulent regimes~\cite{White99}:

\begin{equation}
h_{f}=f\frac{l v^{2}}{2 g d},  
\label{Darcy-Weisbach}
\end{equation}%
where $f$ is called the Darcy friction factor, $d$ is the pipeline diameter, $v$ is the average velocity, and $g$ is the acceleration of gravity. 
Equation (\ref{Darcy-Weisbach}) can be written as a function of the volumetric flow rate $dV/dt=\pi \left( \frac{d}{2}\right) ^{2}v$
 (which is the capacity of the pipeline~\cite{GTE03}), as

\begin{equation}
h_{f}=\frac{16f\,l\left( dV/dt\right) ^{2}}{\pi ^{2}d^{5}}.
\end{equation}
In general, the friction factor $f$ and the pipeline loss $h_{f}$ \ depend
on the pipeline diameter $d$, so the capacity of the pipeline is given by $
c=dV/dt=\frac{\pi }{4}\left( \frac{h_{f}}{f\,l}\right) ^{1/2}d^{5/2}\sim
d^{\gamma }$, where typically $\gamma \simeq 2.6$ for gas pipelines~\cite{GTE03}.


\end{document}